\documentclass[%
superscriptaddress,
reprint,
nofootinbib,
nobibnotes,
aps,
prl,
floatfix]{revtex4-1}
\usepackage{blindtext}
\usepackage{lmodern}
\usepackage{amsmath,bm}
\usepackage{amssymb,amsfonts}
\usepackage{xcolor}
\usepackage{graphicx}
\usepackage{dsfont}

\newcommand{\fjh}[1]{{ #1}}

\begin{document}
\title{Quantum Metrology of Newton's constant with Levitated Mechanical Systems} 
\author{Francis J. Headley}
\email{francis.headley@uni-tuebingen.de}
\affiliation{Institut für Theoretische Physik, Eberhard-Karls-Universität Tübingen, 72076 Tübingen, Germany}

\author{Alessio Belenchia}
\affiliation{Institute of Quantum Technologies, German Aerospace Center (DLR), D-89077 Ulm, Germany}

\author{Mauro Paternostro}
\affiliation{Universita` degli Studi di Palermo, Dipartimento di Fisica e Chimica - Emilio Segre`,
via Archirafi 36, I-90123 Palermo, Italy}

\affiliation{Centre for Quantum Materials and Technologies, School of Mathematics
and Physics, Queen’s University Belfast, BT7 1NN, United Kingdom}

\author{Daniel Braun}
\affiliation{Institut für Theoretische Physik, Eberhard-Karls-Universität Tübingen, 72076 Tübingen, Germany}
\date{\today}
\begin{abstract}
 Newton's constant is the least well-measured among the fundamental constants of Nature, and, indeed, its accurate measurement has long served an experimental challenge. Levitated mechanical systems are attracting growing attention for their promising applications in sensing and as an experimental platform for exploring the intersection between quantum physics and gravitation. Here we propose a mechanical interferometric scheme of interacting levitated oscillators for the accurate estimation of Newton's constant. Our scheme promises to beat the current standard by several orders of magnitude.
 \end{abstract}
\maketitle
\textit{Introduction--}
Accurate measurement of Newton's constant
$G\approx6.67\times10^{-11}\, \text{m}^3\text{kg}^{-1}\text{s}^{-2}$ \cite{CODATA} has long been an experimental challenge, primarily due to the weakness of the gravitational interaction with respect to the other fundamental interactions. To date, $G$ remains the fundamental constant with the highest uncertainty. This uncertainty is due to a host of experiments with large deviations in their experimental findings~\cite{gillies_newtonian_1997,bertoldi_atom_2006,fixler_atom_2007,rosi_precision_2014,newman_measurement_2014,li_measurements_2018}. The precise measurement of $G$ requires tremendous isolation from external sources of noise. This, combined with the difficulty of precisely measuring masses and the impossibility of screening gravitational interactions, leaves the precise value of Newton's constant uncertain. 

Current high-precision measurements of $G$ broadly separate into two approaches. In torsion-balance experiments, one infers $G$ from a direct mechanical response of a macroscopic probe to a source mass, typically via an angular deflection (torque) signal or a time-of-swing modulation \cite{rothleitner2017rsiG,li2018natureG,luo2009prlG}. Atom-interferometric experiments, by contrast, encode gravity into a phase shift of matter waves and extract $G$ from interferometric fringes \cite{rosi2014natureG,lamporesi2008prlG}. Torsion balances benefit from quasi-continuous mechanical readout over long integration times, whereas atom interferometers encode gravity into a phase that is typically estimated from repeated finite-duration ‘single-shot’ runs \cite{cronin2009rmp,bongs2019natrevphys}.

Recently, levitated mechanical systems, i.e.~particles suspended in electromagnetic or optical fields, have emerged as a promising experimental platform to address some of these challenges~\cite{vinante_levitated_2022,prat-camps_ultrasensitive_2017, gonzalez-ballestero_levitodynamics_2021,bose_massive_2023,winkler_steady-state_2025,poddubny_nonequilibrium_2025,weiss_large_2021,rudolph_force-gradient_2022,cosco_enhanced_2021}. Magnetic levitation, in particular, has been shown to offer exceptional isolation from non-gravitational environmental disturbances \cite{gonzalez-ballestero_levitodynamics_2021} as well as long-lived coherent oscillations~\cite{fuchs_magzep_science, timberlake_acceleration_2019, vinante_ultralow_2020}, essential for stable experimental setups capable of  high-precision measurements \cite{ahrens_levitated_2025}. Furthermore, as magnetic levitation allows for stable trapping of Planck-scale masses, it holds the promise to allow for high-precision measurements of the gravitational field at sub-millimetric distances and, in the long run, the detection of the gravitational field of quantum sources~\cite{westphal_measurement_2021, kaltenbaek_macroscopic_2012, belenchia_quantum_2018, weiss_large_2021}. 

 In this work, motivated by the recent advancements in levitomechanics, in particular the cooling of mechanical oscillators close to their quantum ground state \cite{teufel_sideband_2011,chan_laser_2011,abbott_observation_2009, youssefi_squeezed_2023,delic_cooling_2020, wang_ground-state_2024} and the preparation of mechanical systems in squeezed states ~\cite{pirkkalainen_squeezing_2015,wollman_quantum_2015,lecocq_quantum_2015,youssefi_squeezed_2023}, we show how to engineer a mechanical interferometer using two harmonically trapped masses whose gravitational interaction generates a $G$-dependent phase. Although methods for estimating coupling parameters in other optomechanical platforms have received much attention \cite{bernad_optimal_2018,sanavio_fisher-information-based_2020,schneiter_optimal_2020,wang_optomechanical_2025,peng2025pra_optomech_coupling_estimation,ruppert_high-precision_2022,neto_temperature_2022}, we focus instead on a purely gravitational two-body interaction, where $G$ is encoded as a mechanical phase. This yields an interferometric-style experiment, which keeps the sensor as a long-lived mechanical oscillator and the signal as a gravitationally induced phase. We test a variety of input-states and show that this scheme has the potential to estimate $G$ to great accuracy via simple local measurements that saturate the quantum Cram\'er-Rao bound.

\noindent
\textit{Modeling the system--} 
Consider two harmonic oscillators with free Hamiltonian
\begin{equation}
H_0=\sum_{i=1,2} \left(\dfrac{p_i^2}{2m}+\dfrac{1}{2}m\omega_0^2x_i^2\right).
\end{equation}
We assume oscillators with identical masses $m$ and identical trapping frequencies $\omega_0$ that can be adapted e.g.~through the choice of the trap geometry. The non-relativistic gravitational interaction potential between two spherical masses $m$ at distance $r$ between their center of masses reads
\begin{equation}
U(r)=-G\dfrac{m^2}{r}\,, \label{gravpot}
\end{equation}
where $G$ denotes Newton's constant. The assumption of perfect harmonic oscillators is tenable only as far as the oscillating amplitudes $x_1, x_2$ of the particles around their equilibrium positions remain small compared to their distance $d$ at equilibrium and the width of the trap. Writing $r = |d - x_1 + x_2|$ we expand to quadratic order in $\delta x =x_1-x_2\ll d$,
\begin{equation}
H_G=-\dfrac{Gm^2}{d}\left(1+\dfrac{(x_1-x_2)}{d}+\dfrac{(x_1-x_2)^2}{d^2}+\dots\right). \label{hg}
\end{equation}
The full Hamiltonian is
\begin{equation}
    H=H_0+H_G. \label{fullham}
\end{equation}

The linear term in $x_1-x_2$ leads to a $G$-dependent shift of the equilibrium positions of the oscillators relative to the uncoupled case.  It contains information about $G$ but is difficult to exploit, as $G$ cannot be switched off. We therefore focus on the modes with the new, shifted equilibrium positions and denote them as the ``physical modes" $\{x_{1},p_{1},x_{2},p_{2}\}$. In these coordinates, the linear term in \eqref{hg} is eliminated. The treatment including the linear contributions is given in the Appendix. 
We write $x_i,p_i$ in units of the oscillator length, $x_i\to \sqrt{\hbar/m\omega_0}x_i, \ p_i\to \sqrt{\hbar m \omega_0}p_i$. 
After introducing the normal modes $x_\pm=(x_2\pm x_1)/\sqrt{2}$ and $p_\pm=(p_2\pm p_1)/\sqrt{2}$,
the quadratic parts of the Hamiltonian are diagonalized,
\begin{equation}
    H/\hbar=\omega_0a_+^\dagger a_++\omega_0(1-\eta)a_-^\dagger a_-
    -\dfrac{1}{2}\omega_0\eta(a_-^2+a_-^{\dagger2}) \label{ham0},
\end{equation}
with $a_\pm=(x_\pm+ip_\pm)/\sqrt{2}$ and $\eta=2Gm/d^3\omega_0^2$. A dependence on $G$ appears only in the $a_-$ mode, i.e.~the relative motion of the coupled oscillators. Both $\eta$-dependent terms in Eq.~\eqref{ham0} can generate $G$-dependent squeezing of the $a_-$ mode. In the following we investigate the sensitivity of such a system for the estimation of $G$ including the effect of quantum and thermal fluctuations. 
\begin{figure}[t]
\includegraphics[width=0.9\columnwidth]{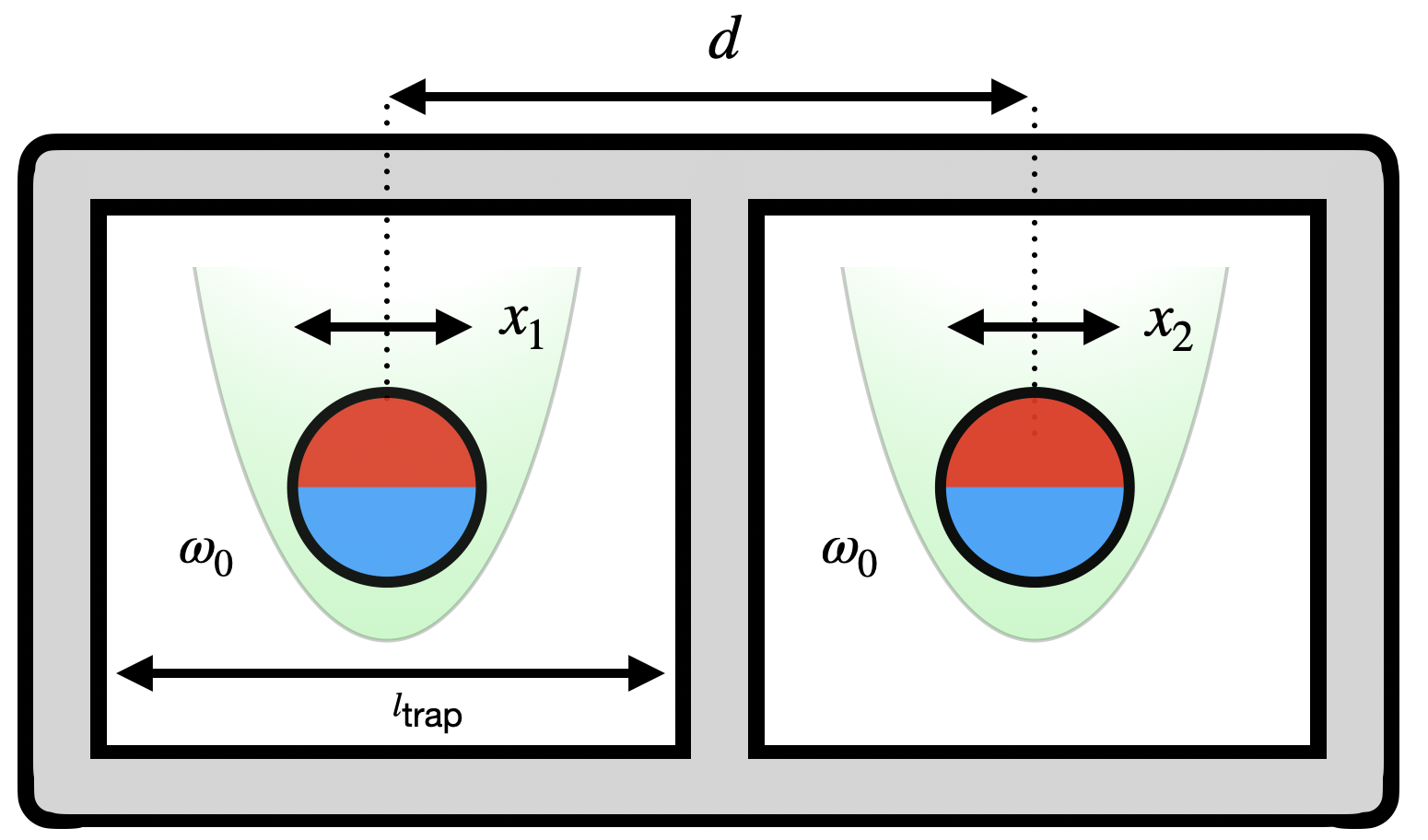}
\caption{Sketch of the set-up. Two objects of mass $m$ separated by a distance $d$ and interacting through gravity levitate in harmonic traps of frequency $\omega_0$, generated e.g.~by trapping small ferromagnets in superconducting boxes. Here $x_{1,2}$ are the displacements from the respective equilibrium positions. 
} \label{fig:diag}
\end{figure}

\noindent 
\textit{Parameter estimation theory --} 
In classical estimation theory, the inverse of the Fisher information (CFI) 
\begin{equation}
F_\text{c}(\theta)=\int_{\mathbb{R}}dxp_\theta(x)\left(\dfrac{\partial\ln p_\theta(x)}{\partial\theta}\right)^2, \label{cfi}
\end{equation}
provides a lower bound to the mean square error of any (unbiased) estimator of a parameter $\theta$ coded in a probability distribution $p_\theta(x)$ of a random variable $x$. This is known as the (classical) Cram\'er-Rao bound (CRB) \cite{Rao1945,Cramer46}. In the quantum mechanical context, $\theta$ parametrizes a density operator $\rho_\theta$ that describes the state of the system, and $x$ is the random outcome of a measurement with $p_\theta(x) = {\rm Tr}[\Pi_x \rho_\theta]$ where  the operator $\Pi_x$ is the element of a positive operator-valued measure (POVM) that formalizes the measurement 
corresponding to the outcome $x$. Optimization over all possible POVMs leads to an upper bound for the classical Fisher information, $F_\text{c}\le F_\text{q}:=\text{tr} \rho_\theta L_\theta^2$, where the symmetric logarithmic derivative $L_\theta$ is defined via $\partial_\theta \rho_\theta=(L_\theta \rho_\theta+ \rho_\theta L_\theta)/2$. The smallest uncertainty (standard deviation) of an unbiased estimator of $\theta$ is then bounded from below by the quantum Cram\'{e}r--Rao bound (QCRB)~\cite{helstrom_quantum_1969,Holevo1982,braunstein_statistical_1994}
\begin{equation}
\delta \theta\geq  \dfrac{1}{\sqrt{QF_\text{c}(\rho_\theta)}}\geq  \dfrac{1}{\sqrt{QF_\text{q}(\rho_\theta)}}\,, 
\end{equation}
where $Q$ is the number of performed experiments. The optimal measurement --- i.e., the one saturating the QCR bound --- is a projective measurement with POVM elements given by the projectors onto the eigenstates of $L_\theta$, but these may not be realizable experimentally. It is therefore common to use the QCRB as a benchmark and check whether experimentally feasible measurements saturate it. 

\noindent
\textit{The interferometric scheme--} We consider as input states separable, Gaussian states of the coupled physical modes $a_{1,2}$ with no initial dependence on $G$. Since the approximate Hamiltonian considered is at most quadratic in the quadratures, Gaussianity is preserved\footnote{This remains the case upon considering noise that can be modelled by a Lindblad master equation with jump operators at most linear in the quadratures.}. After initialising the input states at time $t=0$, the two harmonic oscillators evolve in their traps and interact solely via the Newtonian potential for the duration of the experiment. The transformation from the $a_{1,2}$ modes to the $a_{\pm}$ modes acts as a beamsplitter, and the $a_\pm$ modes serve as the two arms of the interferometer, of which only the $a_-$ arm acquires an additional phase that depends on $G$.  
At a time $t=t_f$ we close the 
interferometer
by performing local projective Gaussian measurements on the physical modes of the oscillators $a_{1,2}$. By  having two freely evolving modes with a phase accumulated between them that carries information on $G$, our  interferometric setup differs from standard measurement schemes where only a single test mass evolves freely, and a source mass is driven~\cite{hoyle_submillimeter_2004, kapner_tests_2007} or displaced with external forces~\cite{fixler_atom_2007,cavendish_xxi_1997}.

\noindent
\textit{Input states--}
We employ the Gaussian formalism to parametrize initial states and solve the dynamics~\cite{ferraro_gaussian_2005, genoni_conditional_2016,pinel_ultimate_2012,  mcmillen_quantum-limited_2017}. Gaussian states are parametrized entirely by the first moments, i.e. the expectation values of the quadratures $\mathbf{r}(t)=(x_1(t),p_1(t),x_2(t),p_2(t))^T$, and covariance matrix $\mathbf{\sigma}$ (CM) where $\sigma_{ij}=\frac{1}{2}\langle\{\hat{r}_i,\hat{r}_j\}\rangle-\langle  \hat{r}_i\rangle\langle \hat{r}_j\rangle$, with $\{\cdot,\cdot\}$ the anti-commutator. 
The relevant parameters which enter into the CM are the squeezing parameters $s_1,s_2$ and thermal occupation number $\bar{n}=\left(\exp[\hbar \omega_0/k_bT]-1\right)^{-1}$ 
\cite{krisnanda_observable_2020} [c.f. Appendices]. 
Squeezed input states can be generated by first exciting the physical modes of the oscillators into coherent states and then modifying
the trap frequencies. Explicitly, a mode of frequency $a_1(\omega_1)$ can be cast as the result of squeezing a different mode of frequency $\omega_0$ according to the relation $a_1(\omega_0)=S^\dagger(s_1)a_1(\omega_1)S(s_1)$, with the squeezing operator  $S(s_1)=\exp\left[\frac{1}{2}(s_1^*a^2_1(\omega_1\right)-s_1a_1^{\dagger2}(\omega_1))]$ and  squeezing parameter $s_1=\frac{1}{2}\log(\omega_0/\omega_1)$~\cite{graham_squeezing_1987}. 
Modifying  
only the frequency of the first mode $a_1(\omega_1)\to a_1(\omega_0)$ and initializing the $a_2(\omega_0)$ mode in the ground state, our input states for zero temperature read 
\begin{equation}
    |\alpha_1\rangle_{a_1(\omega_1)} |0\rangle_{a_2(\omega_0)}=   |\alpha_1,s_1\rangle_{a_1(\omega_0)} |0,0\rangle_{a_2(\omega_0)}\,, \label{state}
\end{equation}
where $|\alpha\rangle_{a_i(\omega_i)}=D(\alpha)_{a_i(\omega)}|0\rangle_{a_i(\omega)}$ is a coherent state of the $a_i(\omega)$-mode, $|\alpha_i,s_i\rangle_{a_i(\omega)}=  
S(s_i)D(\alpha_i)_{a_i(\omega)}|0\rangle_{a_i(\omega)}$ is a squeezed coherent state of the $a_i(\omega)$-mode, and $D(\alpha)_{a_i(\omega)}=\exp[\alpha a_i^\dagger(\omega)-\alpha^* a_i(\omega)]$ is the displacement operator. The amplitudes of the coherent states are calculated through the input first moments [cf. Appendices]. 
{In the following, we consider general-dyne measurements accounted for within the Gaussian formalism~\cite{ferraro_gaussian_2005, genoni_conditional_2016,pinel_ultimate_2012,  mcmillen_quantum-limited_2017}. In particular, we consider both projective and continuous weak local measurements on the physical modes. In the projective case, the CFI is obtained by modifying the covariance matrix  as $\boldsymbol{\sigma}(t) \ \to  \ \boldsymbol{\sigma}(t) + \boldsymbol{\sigma}_m$~\cite{mcmillen_quantum-limited_2017}.  
The  added  term $\boldsymbol{\sigma}_m=\text{Diag}\left[s,s^{-1},s,s^{-1}\right]/2$  corresponds to projective measurements 
of the $a_1(\omega_0), \ a_2(\omega_0)$ modes, with position ($s\to 0$) and momentum ($s\to\infty$) quadratures homodyne detection, and balanced heterodyne detection 
($s=1$). {The Quantum Fisher information for Gaussian states can be expressed as \cite{pinel_ultimate_2012}:
\begin{equation}
F_\text{q}(\rho_\theta) = (\langle\boldsymbol{r}'_\theta\rangle)^T\boldsymbol{\sigma}^{-1}_\theta\langle\boldsymbol{r}_\theta'\rangle+\frac{1}{2(1+P^2)}\text{Tr}\left((\boldsymbol{\sigma}_\theta'\boldsymbol{\sigma}_\theta^{-1})^2\right), \label{ficm}
\end{equation}
where $\bm{A}'=\partial_\theta\bm{A}$ and $P=(\det(\sigma_\theta))^{-1/2}$ is the purity of the state.}
For the case of continuous measurements, indirect homodyne and heterodyne measurements are considered. The environment is assumed to be a Markovian bath with thermal correlations in the input-output formalism. The system interacts with the bath modes, which are then measured. By solving the corresponding stochastic evolution equations for quadratures and CM of the system, we compute the CFI associated to the measured currents, exploiting the Gaussianity of the dynamics~\cite{PhysRevA.95.012116}, as 
\begin{equation}
    F_{c} = \sum_{\text{traj}}\frac{(\partial_\theta p_{\text{traj}})^2}{p_\text{traj}},
\end{equation}
where $p_{\text{traj}}$ is the probability of a certain trajectory that depends on the measurement strategy~\cite{albarelli2018restoring,PRXQuantum.5.020201}. The QFI, and thus the QCR bound, are more challenging to compute for continuously monitored open systems~\cite{yang2025quantum,PhysRevX.13.031012,PhysRevLett.112.170401}. Here we employ an effective QFI which \textit{de facto} gives the ultimate bound for the chosen monitoring strategy~\cite{PhysRevA.93.042121,albarelli2017ultimate,PhysRevA.90.012330} and can be written as the sum of the CFI and the average of the QFI for the conditional state of the system [c.f. Appendices for the details of the calculation].}

\noindent
\textit{Results --} 
We consider a possible experimental realization of the interferometric set-up with realistic parameters: two neodymium magnets with a density of $\varrho=7430\, \text{kg m}^{-3}$, levitated in superconducting traps, see Fig.~\ref{fig:diag}. The magnets are taken to be perfect spheres of radius $r_\text{mass}\approx1.48\,\text{mm}$, corresponding to a mass of $m= 1\times10^{-4}\text{kg}$. The trap size is taken as $l_\text{trap}=4.9\times10^{-3}\,\text{m}$, slightly smaller than the distance between the centers of motion $d=5\times10^{-3}$m. We assume $\omega_0=100\, \text{rad} \text{s}^{-1}$, as  trapping frequencies in the range $\sim10-100\,\text{Hz}$ have been experimentally achieved for levitated neodymium magnets~\cite{vinante_ultralow_2020,fuchs_magzep_science,timberlake_linear_2024}. Squeezed light experiments have yielded high squeezing parameters of up to $s\approx 1.73$, corresponding to 15\,dB~\cite{vahlbruch_observation_2008,vahlbruch_detection_2016}. Such large squeezing may, through the process of state transfer, be imparted onto mechanical systems~\cite{aspelmeyer_cavity_2014}. Therefore,  a squeezing parameter $s\in [-1.8,1.8]$ was used  in our analysis. With these values, we achieve a coupling parameter $\eta=1.068\times10^{-11}$. A superconducting wall between the two magnets shields the electromagnetic forces. {Additional noise sources, including thermal, vacuum, and Casimir effects, are determined to be significantly weaker than the signals analyzed in this study [cf. Appendices].} An experimental realization with similar scales to our set-up with just one levitated particle can be found in Ref.~\cite{vinante_levitated_2022}. 

In order to increase the QFI, one may use large-amplitude input states. However, one must then consider the validity of the truncation of the gravitational potential in Eq.~\eqref{hg} to second order in $\delta x=(x_1-x_2)$. 
To third order in $\delta x$, the relative difference between Eq.~\eqref{hg} and the full expression in Eq.~\eqref{gravpot} is $D_{rel}=(x_1-x_2)^3/d^3$. For $D_{rel}<1\times10^{-3}$, and a maximum amplitude smaller than the center of mass separation, $|x_1-x_2|<\lambda d$, the bound imposes $\lambda<1/10$. For $\lambda=1/10$, with input oscillator amplitude $\langle x(0)\rangle=\lambda l_\text{trap}<\lambda d$ we are well within the appropriate length-scales that justify the truncation in Eq.~\eqref{hg}. Hence, in the following we fix the input amplitude\footnote{For the dimensionful position quadrature.} $\langle x_1(0)\rangle= l_\text{trap}/10$, which implies a dimensionless input amplitude of the $a_1$ mode of $\alpha_1\approx3.374\times10^{12}$ [cf. Appendices]. 
 Introducing larger amplitudes requires considering higher-order terms of the potential and would result in non-linear effects beyond the scopes of this work.

\begin{figure*}[t]
{\bf (a)}\hskip5cm{\bf (b)}\hskip5cm{\bf (c)}\\
\includegraphics[width=0.65\columnwidth]{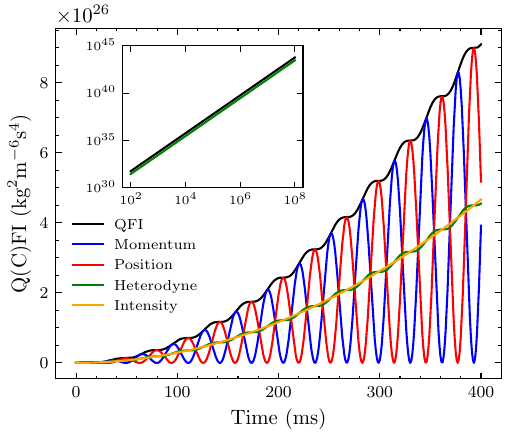}\includegraphics[width=0.65\columnwidth]{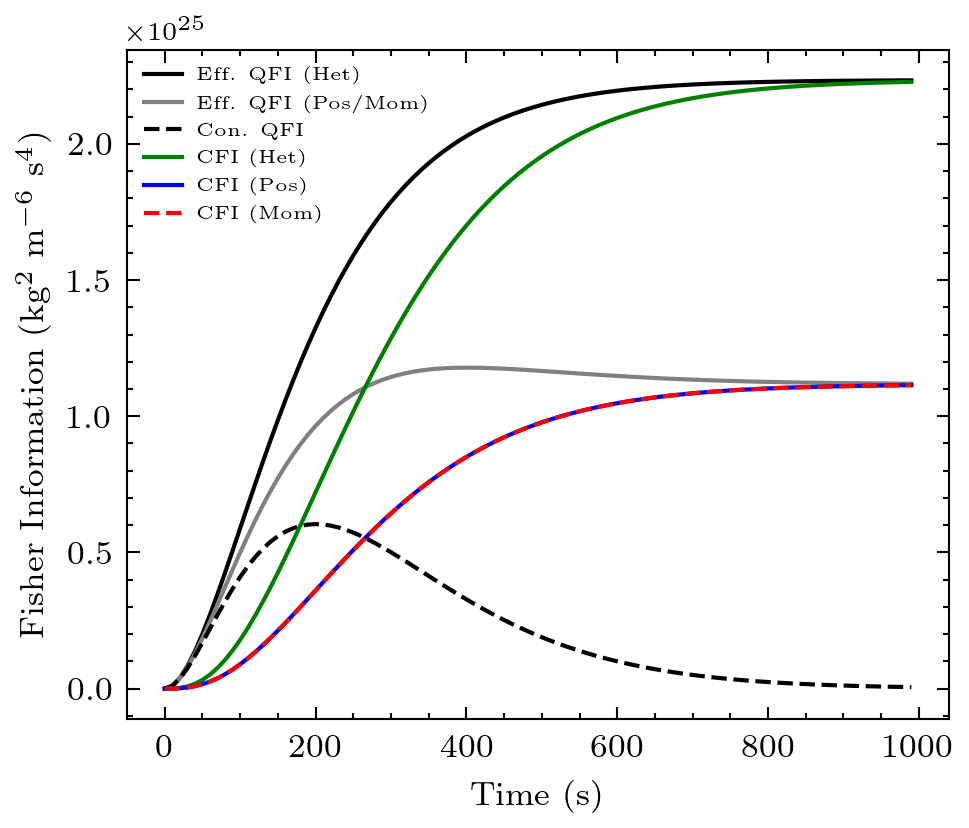}\includegraphics[width=0.65\columnwidth]{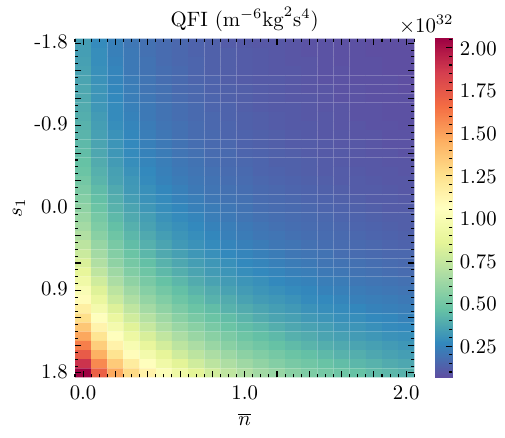} 
\caption{QFI and CFI for different measurement schemes. {\bf (a)} Projective measurements: the black curve is the QFI, calculated using the CM; the red curve represents the CFI for $s\to 0$ (projective position measurement) while the blue curve the CFI for $s \to \infty$ (projective momentum measurement). Finally, the green curve is the CFI for $s \to 1$ (projective heterodyne measurement). We also report in yellow the CFI for intensity measurements. The temperature is $T=0$ ($\bar{n}=0$).{ {\bf (b)} Continuous measurements at finite temperature ($T=1$mK, i.e., $\bar{n}=1.31\times10^6$) and quality factor ($\mathcal{Q}=10^7$). The measurement rate is chosen as
$\Gamma_m=1\times10^{-2}$ with perfect detector efficiency.} Here we report the CFI for the different measurement strategies compared to the corresponding effective QFI [cf. Appendix for details]. The black dashed line represent the effective and conditional QFI of the conditional state of the system. The CFI of the position and momentum measurements are plotted overlapping in blue and red. 
In panels {\bf (a)} and {\bf (b)}, the input state is chosen as the tensor product of two thermal coherent states with parameters $\omega_0=100\ \text{rad}\ \text{s}^{-1}$, $m=1\times10^{-4}$kg, $d=5\times10^{-3}$m,  $\alpha_1\approx3.374\times10^{12}$, $\alpha_2=0$. {\bf (c)} QFI as a function of the squeezing $s_1$ and thermal occupation number $\bar{n}$ for a displaced squeezed thermal state at $t=100$s. 
Other parameters as in panels {\bf (a)}, {where the input amplitude is defined through the initial displacement as $\langle x_i(0)\rangle=\sqrt{2}\alpha_i$.}
} \label{fig:qfist}
\end{figure*}

The results for projective measurements are shown in Fig.~\ref{fig:qfist}~{\bf (a)} and {for continuous measurements in Fig.~\ref{fig:qfist}~{\bf (b)}}. In the former case, measurements of the position and momentum quadratures periodically saturate the QFI, winning out  over heterodyne measurements. For long times, the $t^2$ term in the QFI dominates. With our given set of parameters, we see in Fig.~\ref{fig:qfist} that a projective momentum measurement performed at $t\approx100\,$s saturates the QFI. This corresponds to a sensitivity of $\delta G \approx  1.31\times10^{-16}\, \text{kg}^{-1} \text{m}^3 \text{s}^{-2}$, or relative statistical error $\delta G/G \approx 1.96\times10^{-6}$, beating the current CODATA value of $\delta G/G \approx2.2\times10^{-5}$ \cite{CODATA}  by one order of magnitude. Performing a projective position measurement instead at $t=10^5$\,s, a sensitivity of $\delta G\approx1.31\times 10^{-19}$ is achieved, corresponding to a relative uncertainty of $\delta G/G \approx 1.96\times10^{-9}$, four orders of magnitude better than the CODATA standard uncertainty. Due to the weakness of the gravitational coupling $(Gm/d^3\omega_0^2)\ll 1$, one may approximate the Hamiltonian in Eq.~\eqref{ham0} by neglecting the terms involving $a_i^2$ and  $a_i^{\dagger2}\,(i=1,2)$
\begin{equation}
    H/\hbar\simeq\omega_0a_+^\dagger a_++\omega_0(1-\eta)a_-^\dagger a_-\,.
    \label{hamrwa}
\end{equation}
Coherent, separable initial states $|\alpha_1\rangle_{a_1(\omega_0)}\otimes|\alpha_2\rangle_{a_2(\omega_0)}$ evolving with Eq.~\eqref{hamrwa} then yield a simple analytical expression for the QFI~\cite{muller_pushing_2024}, 
 \begin{equation}     F_\text{q}(G)=4(|\partial_\theta\alpha_1|^2+|\partial_\theta\alpha_2|^2)=\frac{16 m^2}{d^6\omega^2_0}\Delta^2t^2, \label{tsq}
 \end{equation}
where $\Delta=(\alpha_2-\alpha_1)/\sqrt{2}$ parametrizes the 
differences in amplitude 
of the input states.
Eq.~\eqref{tsq} reproduces the leading behavior of the exact result in Fig.~\ref{fig:qfist}\textbf{(a)}.
The $t^2$ scaling of the QFI  
and the corresponding $1/\sqrt{Q}t$ scaling for the QCRB are typical of coherent evolution. For our set of parameters, the prefactor in Eq.~\eqref{tsq} is ${16m^2\Delta^2}/({d^6\omega_0^2})=5.8\times10^{27}\text{kg}^2\, \text{m}^{-6}\ \text{s}^2$, 
and leads to the very rapid increase of the QFI. 
Different from standard interferometric setups in quantum optics, we profit from direct access to the quadratures in the output ports rather than measuring the difference between the two output intensities.  We see in Fig.~\ref{fig:qfist}\textbf{(a)} that
the latter 
also leads to very rapid information gain, albeit only with half the rate compared to homodyne measurements at zero temperature. The CFI for intensity measurements can be calculated using the evolution described by Eq.~\eqref{hamrwa} to be $F_\text{c}^{ph}(G)=F_\text{q}(G)/2$. 
Therefore, such measurements do not saturate the QCRB and are outperformed by homodyne measurements. 

\begin{figure}[b]
\includegraphics[width=0.9\columnwidth]{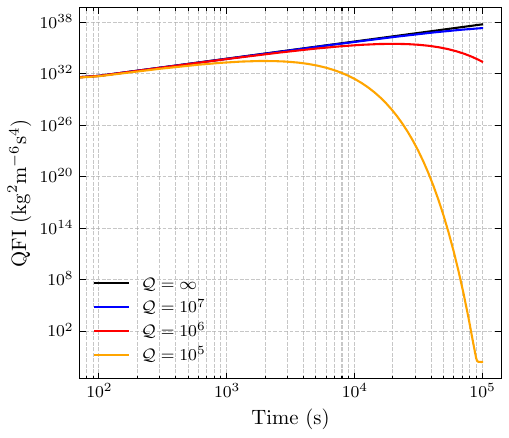}
\caption{QFI against time for different values of the ${\cal Q}$ factor (see the plot legend). All curves shown are for the input state $|\alpha_1,0\rangle_{a_1(\omega_0)}\otimes|0,0\rangle_{a_2(\omega_0)}$ with parameters $\omega_0=100\,\text{rad}\ \text{s}^{-1}$, $m=1\times10^{-4}$\,kg, $d=5\times10^{-3}$\,m,  $\alpha_1\approx3.374\times10^{12}$, $\alpha_2=0$, $\bar{n}=0$.}\label{fig:FILTD}
\end{figure}

\noindent
\textit{Thermal and squeezed states --}
We now consider more general Gaussian input states parametrized by their squeezing $s_1,s_2$, thermal occupation number $\bar{n}$, 
and first and second moments, as outlined before. This relaxes the assumption of ground state preparation in Eq.~\eqref{state}, thus considering more experimentally friendly scenarios. {For thermal and squeezed states the input amplitude is defined through the initial displacement $\langle x_i(0)\rangle=\sqrt{2}\alpha_i$.
Fig.~\ref{fig:qfist} {\bf (c)} shows the impact of squeezing of the $a_1$-mode and the thermal occupation number $\bar{n}$ of both of the input modes on the QFI. Increasing $\bar{n}$ leads to reduced sensitivity at all times. Indeed, the QFI decays with the inverse of the thermal occupation number, $F_\text{q}(G)\propto (1+2\bar{n})^{-1}$, which implies $\delta G \propto (1+2\bar{n})^{1/2}$.  
Fig.~\ref{fig:qfist} {\bf (c)} shows this reduction at $t=100$\,s. Despite this reduction in sensitivity the QCRB remains saturable under projective position and momentum measurements in short and long time frames. {Furthermore, 
for $\bar{n}\gg1$  heterodyne measurements happen to saturate the quantum Cramér-Rao bound  in contrast to the $T=0$ case} [cf. Appendices]. For fixed temperature, squeezing of the $a_1(\omega_0)$'s momentum coordinate ($s_1>0)$ leads to an increase of the value of the QFI at all times. For example, a squeezing parameter of $s_1=1.73$ and thermal occupation $\bar{n}=0$  achieves a relative uncertainty $\delta G/G \approx 1.08\times10^{-6}$ at $t=100s$, approximately a two-fold improvement over the unsqueezed state. Comparable gains are present at all times. In contrast, squeezing the position coordinate ($s_1<0$) decreases the QFI at all times, reducing the achievable sensitivity compared to the unsqueezed case. {When the input state of the $a_1$ mode is chosen in such a way that $\langle p_1(0)\rangle=\sqrt{2}\alpha_1$ and $\langle x_1(0)\rangle=0$, then the behavior is inverted, such that squeezing of the position coordinate $(s_1<0)$ gives a two-fold improvement for all times.}

\noindent
\textit{Effects of damping--} The effect of damping on the sensitivity of the system is considered with respect to currently realistic experimental set-ups, where quality factors ${\cal Q}=\omega/\gamma$ of order ${\cal Q}\approx 10^7$ have been achieved 
\cite{vovrosh_parametric_2017,timberlake_linear_2024,vinante_ultralow_2020,fuchs_magzep_science}. 
Fig.~\ref{fig:FILTD} shows that the QFI reaches a maximum at finite time and decays to very small values for long times limiting the $t^2$ scaling of the QFI to shorter and shorter times for decreasing ${\cal Q}$. 
{Under damping, the QFI scales as $F^\gamma_q(G,\gamma)\approx F_\text{q}(G)e^{-\gamma t}$. {For the tensor product of two displaced thermal states in input, we have $\max F_\text{q}^\gamma(G,\gamma,\bar{n})=(1+2\bar{n})^{-1}(64m^2\Delta^2)/(d^6\omega^2\gamma^2e^2)$ at $t_\text{max}=2/\gamma$.} With our choice of parameters and ${\cal Q}=10^7$, the minimum relative standard uncertainty is $\delta G_\text{min}/G \approx 2.67\times10^{-9}$, gaining four orders of magnitude over the current CODATA value in approximately 2.3 days.} For ${\cal Q}=10^5$ ($\gamma=10^{-3}$\,$\text{s}^{-1}$) the QFI goes to almost zero after approximately $10^5$\,s. The other curves in Fig.~\ref{fig:FILTD}, for which $\gamma\neq0$, show similar behavior at a much longer time scale. For shorter time frames, the effect of damping is minimal. 

 Finally, Fig.~\ref{fig:qfist} {\bf{(b)}} shows the effect of weak, continuous measurements. In this case the system is continuously monitored by measuring (part of) the environment. Details of the computation are reported in the Appendix. We consider a finite temperature $T=1\text{mK}$, a quality factor $\mathcal{Q}=10^7$ and a measurement rate $\Gamma_m=1\times10^{-2}$ with detectors of perfect efficiency~\cite{winkler_steady-state_2025}. We see that all measurement strategies saturate, for long times, the corresponding effective QFI with heterodyne appearing as the favorable strategy for these parameters. { Moreover it can be seen that the measurement strategy had minimal effect on the conditional QFI, as such curves were combined.}

\noindent
\textit{Conclusions--} We have shown that coupled levitated mechanical oscillators offer themselves as a platform for an interferometric measurement of Newton's constant $G$, theoretically beating the current standard uncertainty by four orders of magnitude with a measurement time of about a day. With additional squeezing in one input mode, it is possible to gain further improvements in the sensitivity, also at finite temperature. Although mass estimation and other experimental factors will make it challenging to achieve these levels of sensitivity, our results, together with the rapid progress of experimental implementations, motivate the use of these systems for quantum metrology of small gravitational effects, and, more generally, for the exploration of the boundary between gravitational physics and quantum mechanics.

\noindent
\textit{Acknowledgements --} 
We thank Fabian Müller, Emre Köse, Hendrik Ulbricht, Albert Cabot, Paulo J Paulino and Adriano Braga Barreto for fruitful discussions. We acknowledge the EU EIC Pathfinder project QuCoM (101046973). MP acknowledges financial support from the UK funding agency EPSRC (grant EP/T028424/1), the Royal Society Wolfson Fellowship (RSWF/R3/183013), the Department for the Economy of Northern Ireland under the US-Ireland R\&D Partnership Programme, the ``Italian National Quantum Science and Technology Institute (NQSTI)" (PE0000023) - SPOKE 2 through project ASpEQCt, and the “National Centre for HPC, Big Data and Quantum Computing (HPC)” (CN00000013) – SPOKE 10 through project HyQELM. 

\bibliography{refs} 

\clearpage
\appendix
\section{Input Amplitudes}
For a coherent state, the expectation value of the position and momentum are proportional to the state's amplitude
$ \langle x\rangle=\sqrt{\frac{\hbar}{2m\omega}}(\alpha+\alpha^*) $, $ \langle p\rangle=-i\sqrt{\frac{\hbar m\omega}{2}}(\alpha-\alpha^*)$,
where $x$ and $p$ are dimensionful. As such
\begin{equation}
    \alpha=\sqrt{\dfrac{m\omega}{2\hbar}} \langle x(t)\rangle+i\sqrt{\dfrac{1}{2\hbar m\omega}}\langle p(t)\rangle.
\end{equation}
If we assume at $t=0$ that the expectation value of the momentum $\langle p(0) \rangle=0$, then the input amplitude can be calculated to be
\begin{equation}
    \alpha=\sqrt{\dfrac{m\omega}{2\hbar}} \langle x(0)\rangle.
\end{equation}
Assuming $\omega=100$\,rad\,s$^{-1}$, $m=1\times10^{-4}$\,kg and a trap of size $l_{trap}=4.9\times10^{-3}$\,m with oscillatory amplitude of $l_\text{trap}/10$, then $\alpha\approx3.374\times10^{12}$.
For thermal and squeezed states, the input amplitude is similarly defined through the initial displacement, such that $\langle x_i(0)\rangle=\sqrt{\frac{2\hbar}{m\omega}}\alpha_i$.

\section{States and Evolution}
In this section we lay out the framework of the Gaussian evolution of the system of coupled oscillators. Much is borrowed from~\cite{genoni_conditional_2016}. In continuous variable systems we promote the dimensionless quadratures to operators $\hat{x}_i$ and $\hat{p}_i$, for $i=1,\dots,n$, which if combined to form the vector $\mathbf{\hat{r}}=(\hat{x}_1,\hat{p}_1,\dots,\hat{x}_n,\hat{p}_n)^T$, satisfy the commutation relations:
\begin{equation}
[\hat{r}_i,\hat{r}_j]=i\Omega_{ij},
\end{equation}
where  $i,j=1,\dots,2n$ and $\Omega$ is the symplectic form, usually defined as
\begin{equation}
\Omega=\bigoplus_{j=1}^n\omega, \quad \text{where} \quad \omega=
\begin{pmatrix}
0 & 1 \\
-1 & 0 
\end{pmatrix}.
\end{equation}
The symplectic matrix arises naturally from the commutation relations $[\hat{x}_i,\hat{p}_j]=i\delta_{ij}$. Hamiltonians preserving Gaussianity are at most quadratic in the canonical operators. A general Hamiltonian for Gaussian evolution may be written as
\begin{equation}
\hat{H}=\frac{1}{2}\mathbf{\hat{r}}^TH\mathbf{\hat{r}}+\mathbf{\hat{r}}^T\mathbf{r}_H,
\end{equation}
where $\mathbf{r}_H$ is comprised of linear coefficients.\fjh{ Gaussian states are fully characterized by their first momenta and their covariance matrix (CM) defined as 
\begin{equation}
\sigma_{ij}=\frac{1}{2}\langle\{\hat{r}_i,\hat{r}_j\}\rangle-\langle  \hat{r}_i\rangle\langle \hat{r}_j\rangle,
\end{equation}
with $\{\cdot,\cdot\}$ the anti-commutator. The CM is a positive semi-definite $2n\times 2n$ matrix satisfying the Robertson-Schr\"{o}dinger uncertainty relations
\begin{equation}
\sigma+\frac{i}{2}\Omega\geq 0. \label{uncertainty}
\end{equation}}

\section{Quantum Fisher Information - Covariance Matrix Formulation}

For a probability distribution $p_\theta(x)$ dependent on a parameter $\theta$ which we wish to estimate, the classical Fisher Information (CFI) can be written as:
\begin{equation}
F_\text{c}(\theta)=\int_{\mathbb{R}}dxp_\theta(x)\left(\dfrac{\partial\ln p_\theta(x)}{\partial\theta}\right)^2.
\end{equation}
For a Gaussian quantum state the CFI takes the rather simple form:
\begin{equation}
F_\text{c}(\boldsymbol{r}_\theta,\sigma_{\theta m}) = (\langle\boldsymbol{r}_\theta'\rangle)^T\sigma^{-1}_{\theta m}\langle\boldsymbol{r}_\theta'\rangle+\frac{1}{2}\text{Tr}\left((\sigma_{\theta m}'\sigma_{\theta m}^{-1})^2\right),\label{cfi1}
\end{equation}
where $\sigma_{\theta m}$ is the covariance matrix of the system after performing a measurement. The Quantum Fisher Information (QFI) represents the maximum amount of accessible information about the parameter $\theta$ which can be extracted from the state. The Quantum Fisher information for Gaussian states can be expressed as \cite{pinel_ultimate_2012}:
\begin{equation}
F_\text{q}(\rho_\theta) = (\langle\boldsymbol{r}'_\theta\rangle)^T\sigma^{-1}_\theta\langle\boldsymbol{r}_\theta'\rangle+\frac{1}{2(1+P^2)}\text{Tr}\left((\sigma_\theta'\sigma_\theta^{-1})^2\right), \label{ficm}
\end{equation}
where $\langle\boldsymbol{r}'\rangle$ denotes derivative of the column vector of expectation values for the first moments with respect to the parameter $\theta$ and $\sigma$ is again the covariance matrix. Generally, the best achievable sensitivity of an unbiased estimator of a parameter $\theta$ is bounded from bellow by the Quantum Cram\'{e}r--Rao bound:
\begin{equation}
\delta \theta \geq \delta\theta_{\text{min}} =\dfrac{1}{\sqrt{QF_\text{q}(\rho_\theta)}},
\end{equation}
where Q is the number of performed experiments and $P=(\det(\sigma_\theta))^{-1/2}$ is the purity of the state. Whilst the QCRB gives us the best achievable sensitivity over all unbiased estimators and all POVM measurements, it is important to ask which types of measurements  saturate this bound. 

In the main text, we evaluate the CFI whilst restricting to local Gaussian measurements on both of our modes. An example of such a measurement can be described by a POVM with elements restricted to a pure two-mode Gaussian state with covariance matrix $\sigma_m=\text{diag}[s,s^{-1},s,s^{-1}]/2$ , where $s\in [0,\infty]$ parameterises the degree of squeezing of the elements of the POVM. If such a measurement is performed on a state $\sigma(t)$, the best achievable sensitivity of the measurement is given by the CFI in Eq.~\eqref{cfi1} with the covariance matrix $\sigma(t)+\sigma_m$ \cite{mcmillen_quantum-limited_2017}. 

\section{General Evolution} 
As our full Hamiltonian, Eq. (4) in the main text, is truncated to be at most quadratic in order, \fjh{the resulting unitary dynamics preservers Gaussianity. The same holds for all the noise sources we are considering in our work which are characterized by jump operators in the Lindblad master equation that are at most linear in the quadratures.} Since Gaussian states are fully determined by their first moments and covariance matrix, we are able to determine the full evolution of the system by determining the evolution of these two quantities. 

Here we follow the derivation in Refs.~\cite{genoni_conditional_2016,krisnanda_observable_2020}. We begin by recasting Eq.~(4) in the main text in dimensionless form through the rescaling transformations $x_i\to \sqrt{\hbar/m\omega_0}x_i$ and $p_i\to \sqrt{\hbar m\omega_0}p_i$ resulting in 
\begin{widetext}
\begin{equation}
H/\hbar=\sum_{i=1,2} \dfrac{1}{2}\omega_0\left(p_i^2+\left(1-\dfrac{2Gm}{d^3\omega_0^2}\right)x_i^2\right)-\frac{Gm^2}{\sqrt{d^4\hbar m \omega_0}}x_1+\frac{Gm^2}{\sqrt{d^4\hbar m \omega_0}}x_2+\frac{2Gm}{d^3\omega_0}x_1x_2. \label{bigham}
\end{equation}
\end{widetext}
The Langevin equations of motion of for the dimensionless quadratures read
\begin{align}
\dot{x}_1&=\omega_0 p_1,\nonumber\\
\dot{x}_2&=\omega_0 p_2,\label{eom}\\
\dot{p}_1&=-\omega_0\left(1-\frac{2Gm}{d^3\omega_0^2}\right)x_1-\frac{2Gm}{d^3\omega_0} x_2-\gamma p_1+\nu+\xi_1,\nonumber\\
\dot{p}_2&=-\omega_2\left(1-\frac{2Gm}{d^3\omega_0^2}\right)x_2-\frac{2Gm}{d^3\omega_0} x_1-\gamma p_2-\nu+\xi_2 \nonumber, 
\end{align}
where we have introduced a phenomenological damping rate $\gamma$, Browninan noise terms $\xi_1,\,\xi_2$, and defined $\nu=Gm^2/\sqrt{\hbar m\omega_0 d^4}$. \fjh{We assume a high $\mathcal{Q}$-factor set-up, \( Q = \omega / \gamma \gg 1 \), such that the Brownian noises can be treated as uncolored noise characterized by 
\begin{equation}
    \langle \xi_j(t) \xi_j(t') + \xi_j(t') \xi_j(t) \rangle / 2 \simeq \gamma (2 \bar{n} + 1) \delta(t - t'),
\end{equation}
for \( j = 1, 2 \). }

Eqs.~(\ref{eom}) may be expressed in matrix form as $\dot{\hat{\boldsymbol{r}}}(t)=K\hat{\boldsymbol{r}}(t)+{\bf I}(t)$ with the drift matrix
\begin{equation}
K = 
\begin{pmatrix}
0 & \omega_0 & 0 & 0 \\
\ -\omega_0\left(1-\frac{2Gm}{d^3\omega_0^2}\right) &  -\gamma &\  -\frac{2Gm}{d^3\omega_0} &  0  \\
0 & 0 & 0 & \omega_0\\
-\frac{2Gm}{d^3\omega_0} & 0 & -\omega_0\left(1-\frac{2Gm}{d^3\omega_0^2}\right) &  -\gamma  \
\end{pmatrix},\label{genK}
\end{equation}
and where we have introduced the vector ${\bf I}(t)={\bm v}(t)+{\bm \kappa}$, which is comprised of the noise term ${\bm v}(t)=(0,\xi_1(t),0,\xi_2(t))^T$ and the constant term ${\bm \kappa}=(0,\nu,0,-\nu)^T$.
The solutions to the Langevin equations are given by \cite{krisnanda_observable_2020}
\begin{equation}
\boldsymbol{r}(t)=W_+(t)\boldsymbol{r}(0)+W_+\int_0^tdt' W_-(t'){\bf I}(t'), \label{firstmom}
\end{equation}
where $W_\pm(t)=\exp(\pm Kt)$ are evolution matrices. The dynamics of the covariance matrix is governed by 
\begin{equation}
\dot{\sigma}=K\sigma+\sigma K^T + D, \label{CME}
\end{equation}
where $D=\text{Diag}[0,\gamma(2\bar{n}+1),0,\gamma(2\bar{n}+1)]$is the diffusion matrix describing the coupling of the system to the environment~\cite{genoni_conditional_2016}, $\bar{n}=(\exp(\hbar\omega/k_BT)-1)^{-1}$ is the thermal occupation number, and $T$ the temperature of the environment. The solutions to Eq.~\eqref{CME} are of the form
\begin{align}
\sigma(t)=&\ W_+(t)\sigma(0)W_+^T(t)\nonumber\\
&+W_+(t)\int_0^tdt'W_-(t')DW_-^T(t')DW^T_+(t),
\end{align}
which, upon integration, gives us the linear equation
\begin{align}
K{\bm\sigma}(t)+{\bm\sigma}(t)K^T=&-D+KW_+(t){\bm\sigma}(0)W_+^T(t)\nonumber\\
&+W_+(t){\bm\sigma}(0)W_+^T(t)K^T\label{linear}\\
&+W_+(t)DW_+(t)^T. \nonumber
\end{align}

\section{Initial State Parameterisation} 
We consider an uncorrelated input state of the oscillators. In this case, the covariance matrix takes the block-diagonal form
\begin{equation}
{\bm\sigma}(0)=
\begin{pmatrix}
\ {\bm\sigma}_1 & 0 \\ 
0 & {\bm\sigma}_2 \
\end{pmatrix}, \label{init}
\end{equation}
where ${\bm\sigma}_{1(2)}$ is the covariance matrix for the first (second) oscillator respectively. Through local symplectic transformations, each of them can be cast in the form
\begin{equation}
{\bm\sigma}_{i}(\bar{n},s) =
\begin{pmatrix}
\ \frac{1}{2}(2\bar{n}_i+1)e^{s_i} \ & \ 0 \ \\ 
\ 0 \ & \ \frac{1}{2}(2\bar{n}_i+1)e^{-s_i} \
\end{pmatrix}, \label{param}
\end{equation}
which is paparameterized through the local thermal excitation number $\bar{n}_i$ and degree of squeezing $s_i$~(i=1,2).

\section{Quantum Fisher Information from Coherent Amplitudes} \label{AMsection}
 
We note that Eq.(3) in the main text contains the term $\frac{Gm^2}{d^2}(x_1-x_2)$, which displaces the ground state creating a coherent state that evolves under the remainder of the Hamiltonian. For completeness, we include here the contribution of this displacement in the derivation of the QFI. A suitable change of coordinates allows us to discard such displacement. This is the approach taken in the main text, addressing the situation where both oscillators are in equilibrium positions, including the action of gravity, and then one oscillator is excited. However, one might also envisage situations where one  has access to the shift of equilibrium positions due to gravity, e.g.~by bringing the mass in from positions where they are far apart. As the Hamiltonian is at most quadratic, it maps coherent states to coherent states. Thus, we can calculate the QFI through the amplitudes of the aforementioned induced coherent states. We begin by translating our Hamiltonian into the coordinate basis of the normal modes $ x_\pm=\frac{1}{\sqrt{2}}(x_2\pm x_1),$ $p_\pm=\frac{1}{\sqrt{2}}(p_2\pm p_1).$ The position and momentum operators may be expressed as the sum of these operators as 
\begin{equation}
    x_\pm=\frac{1}{\sqrt{2}}(a_\pm+a^\dagger_\pm),\quad p_\pm=-\frac{i}{\sqrt{2}}(a_\pm-a^\dagger_\pm). \label{qta}
\end{equation}
In the normal basis, the quadratic terms of the Hamiltonian diagonalise, with only a linear term remaining:
\begin{equation}
    H/\hbar=\frac{1}{2}\omega_0(p_+^2+ x_+^2+p_-^2+(1-2\eta)x_-^2)+\sqrt{2}\nu x_-.
\end{equation}
Recasting again in terms of the creation and annihilation operators of the normal modes in Eq.~(\ref{qta}), we get
\begin{align}
    H/\hbar=\frac{1}{2}\omega_0{\Big[}&(a_+^\dagger a_++a_+a_+^\dagger)+(1-\eta)(a_-^\dagger a_-+a_-a_-^\dagger)\nonumber\\
    & -\eta(a_-^2+(a_-^\dagger)^2){\Big]}+\nu(a_-+a_-^\dagger).
\end{align}
\textit{Rotating Wave Approximation --} Due to the weakness of the gravitational coupling $Gm/(d^3\omega^2)\ll1$ we are able to make an approximation in which the  $a^2,\ (a^\dagger)^2$ terms are neglected (sometimes called "rotating wave approximation" (RWA), even though here no resonance phenomenon is involved): Transforming into the corotating frame, we see that the $a^2,\ (a^\dagger)^2$ terms are fast oscillating and, as such, can be disregarded. This is also justified a posteriori by a comparison of the exact QFI and the QFI using the Hamiltonian in Eq.(9) in the main text, this comparison can be seen in Fig.~\ref{fig:RWA}. 
\begin{figure}[t]
\includegraphics[width=\columnwidth]{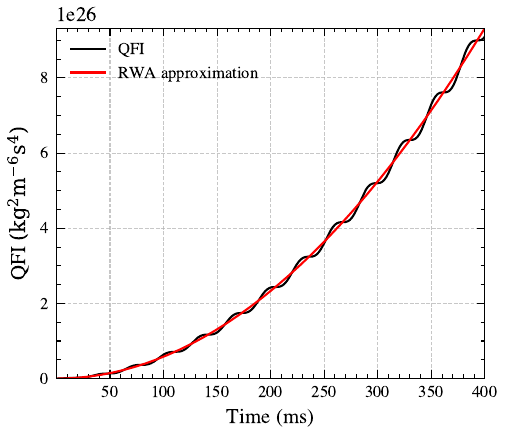}
\caption{Comparison of the exact QFI (Black) and the approximation  Eq.(10) in the main text, which utilizes the corotating frame for input state $|\alpha_1,0\rangle_{a_1(\omega_0)}\otimes|0,0\rangle_{a_2(\omega_0)}$ with parameters $\omega_0=100\,\text{rad}\ \text{s}^{-1}$, $m=1\times10^{-4}$\,kg, $d=5\times10^{-3}$\,m,  $\alpha_1\approx3.374\times10^{12}$, $\alpha_2=0$, $\bar{n}=0$.}\label{fig:RWA}
\end{figure}
The resulting Hamiltonian reads
\begin{equation}
    H/\hbar\approx\omega_0a_+^\dagger a_++\omega_0(1-\eta)a_-^\dagger a_-+\nu(a_-+a_-^\dagger). \label{ham01}
\end{equation}

In order to deal with the linear term, we redefine our $a_-$ mode via a linear translation, sending $a_-\to b\equiv a_-+\kappa$, where $\kappa=\nu/(\omega_0(1-\eta))$. The Hamiltonian becomes completely diagonal:
   $ H/\hbar=\omega_0a_+^\dagger a_++\omega_0(1-\eta)b^\dagger b.$
We may easily translate between coherent states of the $a_{1/2}$ modes and the $a_\pm$ modes
\begin{align}
    |\alpha \rangle_{a_1}| \beta \rangle_{a_2}&=|(\alpha+\beta)/\sqrt{2}\rangle_{a_+}|(\beta-\alpha)/\sqrt{2}\rangle_{a_-},\\
    |\mu \rangle_{a_+}| \nu \rangle_{a_-}&=|(\mu-\nu)/\sqrt{2}\rangle_{a_1}|(\mu+\nu)/\sqrt{2}\rangle_{a_2},
\end{align}
where the lower index refers to the operator which defines the ground state. The ground state of $a_-$ is a coherent state of the $b$ mode with amplitude $\kappa$. Explicitly, we have $|0\rangle_{a_-}=|\kappa\rangle_b\quad \text{and}\quad  |0\rangle_b=|-\kappa\rangle_{a_-}.$ Furthermore we may translate coherent states of $a_-$ into their corresponding coherent state in $b$, $|\alpha\rangle_{a_-}\approx|\alpha+\kappa\rangle_{b}$, 
where we have discarded the phase factor. The state evolves correspondingly as
\begin{equation}
\begin{aligned}
e^{-i H t/\hbar}|\alpha\rangle_{a_1}|\beta\rangle_{a_2}
    &= \left|e^{-i\omega_0t}\frac{\alpha+\beta}{\sqrt{2}}\right\rangle_{a_+}\\&\otimes
    \left|e^{-i\omega_0(1-\eta)t}\left(\frac{\beta-\alpha}{\sqrt{2}}+\kappa\right)-\kappa\right\rangle_{a_-}. \label{flip}
\end{aligned}
\end{equation}
For coherent states, the QFI is given to be the absolute value squared of the derivative of the amplitude with respect to the parameter we wish to measure. For the product of two coherent states $|\alpha\rangle\otimes|\beta\rangle$ the QFI is $F_\text{q}(\theta)=4(|\partial_\theta\alpha|^2+|\partial_\theta\beta|^2)$. The $G$ dependence of the state in Eq.~\eqref{flip} is coded entirely in the $a_-$ mode. Inserting the amplitudes of Eq.~\eqref{flip} into the expression of the QFI for a coherent state
\begin{equation}
     F_\text{q}(G)=4\left|\partial_G\left(e^{-i\omega_0(1-\eta)t}\left(\Delta+\kappa\right)-\kappa\right)\right|^2, 
\end{equation}
where $\Delta=(\beta-\alpha)/\sqrt{2}$ parameterises the amplitude of the input state of the $a_-$ mode. For longer time scales the $t^2$ terms come to dominate the dynamics, leading to a divergence of the QFI for $t\to\infty$ and a $1/\sqrt{N}t$ scaling for the Quantum Cram\'{e}r--Rao bound, 
    \begin{equation}
        F_\text{q}(G)=\dots+16\left(\frac{ G m^{5/2}}{d^2 \sqrt{\omega_0 \hbar}
    \left(d^3 \omega_0^2-2 G m\right)}+\frac{ m}{d^3\omega_0}\Delta\right)^2t^2. 
    \end{equation}
where we have used $\eta=2Gm/d^3\omega_0^2$, $\kappa=\nu/(\omega_0(1-\eta))$ and $\nu=Gm^2/\sqrt{\hbar m\omega_0 d^4}$. If we do not consider the shift in the equilibrium position, then this further reduces to 
 \begin{equation}
     F_\text{q}(G)=16\left(\frac{ m}{d^3\omega_0}\Delta\right)^2t^2. 
 \end{equation}

\vspace{2mm}
\section{CFI of Intensity Measurement.}
For a coherent state $|\beta\rangle$, the CFI associated with an intensity 
measurement is~\cite{muller_pushing_2024} 
\begin{equation}
    F_{int}(\theta)=4\text{Re}\left[\dfrac{\beta^*}{|\beta|}\dfrac{\partial\beta}{\partial\theta}\right]^2\,.
\end{equation}
In quantum optics the intensity measurement would correspond to photon counting.  For the mechanical oscillators considered here, it can be achieved by measuring the energy of the oscillators, which in turn can be achieved by measuring the amplitude $x_\text{max}$ and frequency of the oscillations, knowing the mass of the oscillators. From this one gets the potential energy $m\omega^2 x_\text{max}^2/2$ at times when the kinetic energy vanishes and hence total energy equals the potential energy. For a product state of $m$ coherent states $|\Phi\rangle=|\beta_1\rangle\otimes\dots\otimes|\beta_m\rangle$,  the CFI is the sum of the CFI of each mode
\begin{equation}
    F_\text{c}^{ph}(\theta)=4\sum_{i=1}^m\left(\text{Re}\left[\dfrac{\beta_i^*}{|\beta_i|}\dfrac{\partial\beta_i}{\partial\theta}\right]\right)^2. \label{cfi:ph}
\end{equation}
To prove this, we use 
    $p_\theta(n)=|\langle n|\beta\rangle|^2=e^{-|\beta_\theta|^2}\dfrac{(|\beta_\theta|^2)^n}{n!}$
to get 
\begin{equation}
    \dfrac{\partial p_\theta(n)}{\partial\theta}=-e^{-|\beta_\theta|^2}\dfrac{\partial |\beta|^2}{\partial\theta}\dfrac{(|\beta_\theta|^2)^n}{n!}+e^{-|\beta_\theta|^2}\dfrac{\partial |\beta|^2}{\partial\theta}\dfrac{(|\beta_\theta|^2)^{n-1}}{(n-1)!}
\end{equation}
The calculation of the CFI for the photon resolving measurement on a single mode leads to
\begin{widetext}
    \begin{align}
        F_\text{c}^{ph}(\theta)&=\sum_n\dfrac{1}{p_\theta(n)}\left(\dfrac{\partial p_\theta(n)}{\partial\theta}\right)^2
        =\sum_ne^{-|\beta_\theta|^2}\left(\dfrac{\partial |\beta_\theta|^2}{\partial\theta}\right)^2\left(\dfrac{|\beta_\theta|^{2n}}{n!}-2\dfrac{|\beta_\theta|^{2(n-1)}}{(n-1)!}+\dfrac{n!(|\beta_\theta|^{2(n-2)}}{(n-1)!^2}\right)\nonumber\\
        &=e^{-|\beta_\theta|^2}\left(\dfrac{\partial |\beta_\theta|^2}{\partial\theta}\right)^2\left[e^{|\beta_\theta|^2}-2e^{|\beta_\theta|^2}+e^{|\beta_\theta|^2}\left(1+\dfrac{1}{|\beta_\theta|^2}\right)\right]
        =4\text{Re}\left[\dfrac{\beta^*}{|\beta|}\dfrac{\partial\beta}{\partial\theta}\right]^2.
    \end{align}
\end{widetext}
This can then be extended in a straightforward manner to a product state of $m$ modes~\cite{muller_pushing_2024}. By disregarding the constant shift $\kappa$, we express Eq.~\eqref{flip} in terms of the physical modes $a_1$ and  $a_2$, and calculate the CFI for intensity measurements on them. After a simple calculation one arrives at the approximate expression for the CFI relating to intensity measurements,
\begin{equation}
    F_{c}^{ph}(G)=8\left(\frac{ m}{d^3\omega_0}\Delta\right)^2t^2.
\end{equation}
 For our system, the output physical modes of the interferometer,  remain in a product state of coherent states,
    $|\alpha(1+e^{-i\omega_0(1-\eta)t})/2\rangle_{a_1}|\alpha(1+e^{-i\omega_0(1-\eta)t})/2\rangle_{a_2}$
which follows from Eq.~\eqref{flip} (modulo shift of $\kappa$ in the equilibrium positions). Straightforward calculations using Eq.~\eqref{cfi:ph} yield
    $F_\text{c}^{ph}(G)=8\frac{ m^2}{d^6\omega_0^2}\Delta^2t^2$.

\fjh{\section{Thermal Displacement Effects}

\subsection{Particle Thermal Displacement}
For two levitated magnets with mass $m = 10^{-4} \, \text{kg}$, trap frequency $\omega_0 = 100 \, \text{rad/s}$, and temperature $T = 1 \, \text{mK}$, the root-mean-square (RMS) thermal displacement is calculated as:
\begin{equation}
\delta x_{\text{thermal}} = \sqrt{\frac{k_B T}{m \omega_0^2}} \approx 1.18 \times 10^{-13} \, \text{m},
\end{equation}
where $k_B$ is the Boltzmann constant. For two particles separated by a distance $d = 5 \, \text{mm}$, the relative distance fluctuation is:
\begin{equation}
\delta d = \sqrt{2} \cdot \delta x_{\text{thermal}} \approx 1.67 \times 10^{-13} \, \text{m}
\end{equation}
The relative error in the gravitational coupling parameter $\eta = \frac{2 G m}{d^3 \omega_0^2}$ is:
\begin{equation}
\frac{\delta \eta}{\eta} = 3 \frac{\delta d}{d} \approx 1 \times 10^{-10}.
\end{equation}
This is rather low compared the target sensitivity of $\delta G / G \approx 1.96 \times 10^{-9}$ and is negligible compared to the achievable sensitivity of $\delta G / G \approx 4.33 \times 10^{-6}$ at current experimental temperatures and quality factors. The impact of thermal fluctuations diminishes at lower temperatures due to the $\sqrt{T}$ scaling.

\subsection{Faraday Shield Thermal Displacement}
For a Faraday shield made of lead with a thickness of $t_\text{trap}=(d-l_\text{trap})=0.1\times10^{-3} \, \text{m}$, length of $l=1\times10^{-2} \, \text{m}$, width of $w=2\times10^{-2}\, \text{m}$ and a Young's 
modulus 
of $E=13\, \text{GPa}$, the stiffness of the shield can be approximated as $k_{\text{shield}}=Ewt_\text{trap}^3/4l^3 \approx 65 \, \text{N} \, \text{m}^{-1}$ \cite{poggi_method_2005}. At $T = 1 \, \text{mK}$, the RMS thermal displacement is:
\begin{equation}
\delta z = \sqrt{\frac{k_B T}{k_{\text{shield}}}} \approx 1.46 \times 10^{-14} \, \text{m}.
\end{equation}
The relative distance error due to the shield's displacement is:
\begin{equation}
\frac{\delta d}{d} \approx \frac{\delta z}{d} \approx \frac{1.46 \times 10^{-14}}{5 \times 10^{-3}} \approx 2.91 \times 10^{-12}.
\end{equation}
This contribution is also negligible compared to the target sensitivity of $\delta G / G \approx 1.96 \times 10^{-9}$, confirming that thermal displacements of the Faraday shield do not significantly impact the precision of the gravitational constant measurement.

\section{Casimir and Vacuum Effects}
\subsection{Vacuum effects}
In magnetic levitation within a superconducting cavity, the levitation mechanism is passive, eliminating the need for optical or microwave cavities and active fields. However the models considered above also apply to optically levitated set-ups. 
As in the main text, 
the gravitational interaction yields an effective coupling parameter $\eta \sim 10^{-11}$. In contrast, vacuum-induced effects, such as single-photon radiation-pressure coupling, scale as $\hbar \omega_c (x_{\mathrm{zpf}}/L)$ \cite{aspelmeyer_cavity_2014, butera_corrections_2025, sala_exploring_2018}. For the parameters in the main text ($m = 10^{-4}$\,kg, $\omega_0 = 100$\,rad/s), the zero-point fluctuation amplitude is $x_{\mathrm{zpf}} \approx 7.26 \times 10^{-17}$\,m. With a trap (cavity) scale $L \sim 5$\,mm, the ratio $(x_{\mathrm{zpf}}/L) \sim 10^{-14}$, making vacuum-induced effects negligible by over three orders of magnitude. To quadratic order, this scales as $(x_{\mathrm{zpf}}/L)^2 \sim 10^{-28}$, further confirming their insignificance.
\subsection{Casimir effects}
When considering the Casimir forces between the levitated oscillators and the plates, the proximity-force approximation of the plane-sphere Casimir energy $\mathcal{E}_{PS}^{PFS}$ \cite{maia_neto_casimir_2008} is given 
\begin{equation}
    \mathcal{E}_{PS}^{PFS}=-\dfrac{\hbar c \pi^3r}{720(l/2-r)^2}
\end{equation}
where $r$ is the radius of the oscillator, and $l$ the distance between the two plates. Taking $l=4.9$mm and $r\approx 1.5$mm, we see that the ratio between the Casimir and gravitational term is $\sim1\times10^{-8}$. It is thus legitimate to ignore the Casimir terms in our model.

\section{Continuous measurement}
\begin{figure*}[t]
\includegraphics[width=0.65\columnwidth]{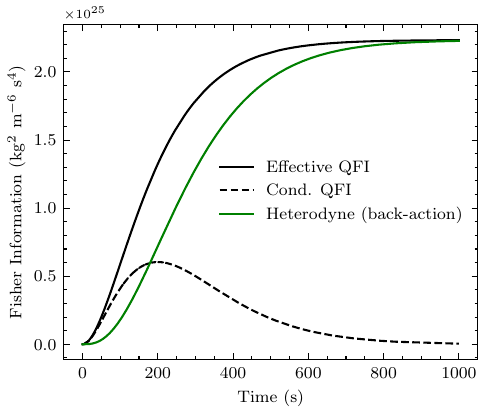}\includegraphics[width=0.65\columnwidth]{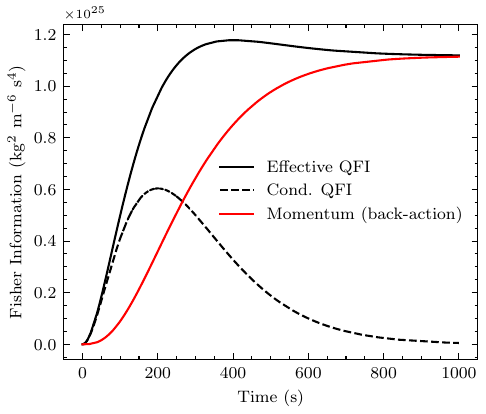}\includegraphics[width=0.65\columnwidth]{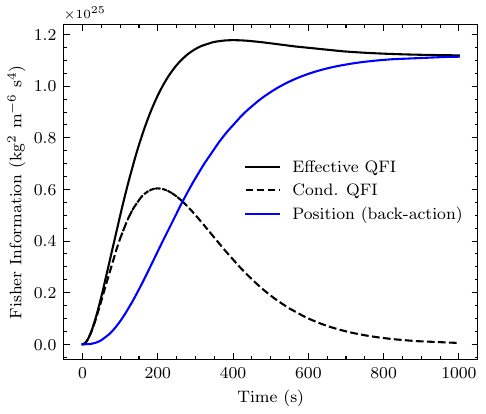} 
\caption{\fjh{Here we compare the left-hand side of eq.~\eqref{effQFI}, representing the effective QFI (black curves) with the second term on the right-hand side of eq.~\eqref{effQFI} --- representing the conditional QFI (dashed black lines) --- and the first term on the right-hand side of the same equation that is the CFI (colored curves). The three panels show these quantities for the three measurement strategies considered. The parameters are as in the main text with $\bar{n}=1.31\times 10^{6}$ ($T=1\,\text{mK}$) and $\Gamma_m=1\times10^{-2}$.}}
 \label{fig:condqfi}
\end{figure*}

Here we cover the details of the dynamics under continuous measurements. 

First, note that typically, for systems involving levitated magnets, detection is done using SQUID-loop detectors, which provide a continuous homodyne signal rather than a projective measurement. This  motivates the need to investigate continuously monitored systems. We do this by considering continuous general-dyne measurements. These make the dynamics stochastic whilst preserving Gaussianity, allowing us to keep using the Gaussian formalism, albeit with some important modification, instead of resorting to stochastic master equations. 

Continuous measurements are ubiquitous in laboratories and levitated set-ups. When a system is continuously measured, the (classical) information about the parameter of interest is contained in the continuous stochastic signals recorded by the detectors. 
However, the QFI requires the knowledge of the full system-environment quantum state to be properly estimated. This is in general not an easy task and also the subject of current investigations~\cite{yang2025quantum,PhysRevX.13.031012,PhysRevLett.112.170401}. Nonetheless, effective QFIs can be readily estimated given the stochastic dynamics of the system as discussed in~\cite{albarelli2018restoring}. Below we follow the latter route ~\cite{serafini2023quantum,genoni_conditional_2016,PhysRevA.95.012116}.

Let us consider a Gaussian system coupled to a large Markovian environment. The environment is described by incoming modes $\hat{\mathbf{r}}_b(t)$, each interacting with the system at a given time $t$ in the input-output formalism. 
The correlations of these modes are characterized by
\begin{equation}
\langle \{\hat{\mathbf{r}}_b(t), \hat{\mathbf{r}}_b^\dagger(t')\} \rangle =\sigma_b \delta(t - t'), \quad \sigma_b + \frac{i}{2}\Omega \geq 0,
\end{equation}
where $\sigma_b$ is the environment's covariance matrix. The coupling between the system and (measured) environment in an interval $dt$ is given by
\begin{equation}
\hat{\mathcal{H}}_C dt  = \hat{\mathbf{r}}^T C \hat{\mathbf{r}}_b(t) dt. 
\end{equation}
Given this interaction, and considering the continuous monitoring of the bath modes via general-dyne measurements characterized via the CM $\sigma_m$, stochastic dynamical equations for the evolution of the first moments and the CM of the system can be derived [cf.~\cite{serafini2023quantum,genoni_conditional_2016} for the details of the derivation]. We have
\begin{align}
    & d\boldsymbol{r}(t)=K\boldsymbol{r}(t)dt+\left(\dfrac{\sigma B+N}{\sqrt{{2}}}\right)\boldsymbol{dw}\\
    & \dfrac{d\sigma}{dt}=K\sigma+\sigma K^T+D+(\sigma B+N)(\sigma B+N)^T,
\end{align}
where $B=C\Omega^T(\sigma_b+\sigma_m)^{-1/2}$, $N=\Omega C\sigma_b(\sigma_b+\sigma_m)^{-1/2}$, and $\boldsymbol{dw}$ is a vector of independent Wiener increments $dw_i dw_j = \delta_{ij}dt$.

As discussed in~\cite{PhysRevA.95.012116}, the outcomes of measuring the bath mode are
distributed according to a Gaussian multi-variate distribution with mean value $\boldsymbol{x}_m = \Omega C^T \boldsymbol{r}(t) dW$, with $dW$ a real Wiener increment, and covariance matrix $(\sigma_b + \sigma_m)$. This allows the author in~\cite{PhysRevA.95.012116} to derive the infinitesimal CFI for a specific trajectory $dF_t^{(\text{traj})}$.
\begin{equation}
    dF_t^{(\text{traj})}={2}\langle{\mathbf{r}}'\rangle^TC\Omega^T(\sigma_b+\sigma_m)^{-1}\Omega C^T\langle{\mathbf{r}}'\rangle dt.
\end{equation}
This is a stochastic quantity. Thus the CFI is computed by averaging it over the trajectories 
\begin{equation}
    dF_\text{c}(p_\text{traj})=\mathbb{E}_{dw}[dF^{(\text{traj})}_t],
\end{equation}
and integrating the result in time:
\begin{equation}
    F_\text{c}(p)=\int_0^tdF_\text{c}(p_\text{traj}).
\end{equation}
Finally, the ultimate bound on the FI given a specific continuous measurement estimation strategy is discussed in~\cite{albarelli2018restoring} as given by
\begin{equation}\label{effQFI}
    F_{\text{EffQ}} = F_\text{c}(p_{\text{traj}})+\sum_{\text{traj}}p_\text{traj}F_\text{q}(\rho_\text{c}).
\end{equation}
Here, the first term is the CFI computed as just described. The second term is the average over the stochastic trajectories of the QFI of the conditional state of the system solution of the stochastic master equation. We use this last expression in order to identify the ultimate bound for the general-dyne measurement strategies that we consider in the main text.}



In the figures, we consider $\sigma_b=\frac{1}{2}(2\bar{n}+1)\mathbb{I}_4$ and $C=\sqrt{\Gamma_m}\mathbb{I}_{4}$. In the same vein as~\cite{winkler_steady-state_2025}, we take $\Gamma_m=1\times10^{-2}$ and fix $\gamma=1\times 10^{-5}$. 
Figures~\ref{fig:condqfi} compare the effective QFI and CFI for different measurement strategies. We see that, for the values of the parameters chosen, the heterodyne strategy results are favorable. Furthermore, we see how all the strategies in the long time saturate the corresponding effective QFI bound.


\end{document}